# Inversion of oxygen potential transitions at grain boundaries of SOFC/SOEC electrolytes

Yanhao Dong[1] (dongyh@mit.edu), I-Wei Chen[2]

[1]*Department of Nuclear Science and Engineering, Massachusetts Institute of Technology, Cambridge, MA 02139, USA*

[2]*Department of Material Science and Engineering, University of Pennsylvania, Philadelphia, PA 19104, USA*

## Abstract

Solid oxide fuel/electrolyzer cell (SOFC/SOEC) converts energy between chemical and electrical forms inversely. Yet electrolyte degradation takes place much more severe for SOEC than SOFC during long-term operations. By solving transport equations, we found very large oxygen potential gradients and sharp oxygen potential transitions at grain boundaries of polycrystalline SOFC/SOEC electrolyte. Surprisingly, an inversion of oxygen potential transitions was identified, suggesting a fundamentally different transport mechanism for minor electronic charge carriers. Such findings could be critical to understand and eliminate SOFC/SOEC degradations in practical applications.

## I. Introduction

Solid oxide electrochemical cell (SOC)[1-3] converts energy between fuel and electricity and has been considered as a key component of the future energy economy. It is typically operated between 600 to 1000 ºC, which enables flexible fuel selections and high efficiency. On the other hand, due to the high operation temperatures and slow heat-up cycles, continuous operation over thousands of hours is preferred, which places a strict requirement on its long-term stability. Therefore, understanding and eliminating the degradation are of great interests[4-13] and we shall focus on the electrolyte part in this work, which has been argued to be the main cause of continuous impedance increase during cell operation[4]. The electrolyte of SOC is a dense ceramic layer that conducts oxygen ions, with yttria stabilized zirconia (YSZ) being the most popular one. While most efforts have been on taken to improve ionic conductivity and minimize ohmic loss across the electrolyte, it has been pointed out that the minor electron/hole conduction cannot be neglected as long as local equilibrium is considered so it is of equal practical importance[4-8]. Specifically, electron/hole conduction is a necessity to determine the chemical potential of molecular oxygen (oxygen potential in short hereafter) as well as its spatial distribution inside the electrolyte. The local oxygen potential would then define the thermodynamics and affect various material properties, including concentrations and conductivities of electrons and holes[14], phase stability[15], mechanical properties[16], chemical expansion and stress[17,18], microstructural evolution[19-21], pore/oxygen bubble formation[4] and ultimately degradation/stability of the electrolyte. The solution of oxygen potential distribution is based on transport equations[7,8,22,23] and previous treatments all assumed a homogeneous "effective medium" whose transport properties solely depend on local oxygen potential, analogous to a single-crystalline electrolyte in some sense. Yet, all electrolyte layers are polycrystalline, sintered from ceramic powders and it is well known that grain boundaries (or more strictly speaking, space charge layers extending several nanometers from grain boundary cores) are blocking to oxygen ions[24,25] hence have distinct transport properties with respect to the grain interiors. Its effect on oxygen potential distribution inside the electrolyte is not known to any extend, which shall be the theme of the present study.

In practical applications, SOC can be operated reversibly, either as a solid oxide fuel cell (SOFC) utilizing chemical fuels for power generation or as a solid oxide electrolyzer cell (SOEC) using electricity to produce fuels. Faster degradation rate has been identified for SOEC than SOFC, leading to observable pore/oxygen bubble generations and line-ups along grain boundaries inside the electrolyte[4,9-11]. The cause has been attributed to larger operational current densities and higher oxygen partial pressures on the anode side of SOEC. However, it cannot be the whole story because the above reasons fail to explain (i) why reversible operations between SOFC and SOEC modes hugely eliminate degradation[4] and (ii) why pores/bubbles preferentially form at the grain boundaries perpendicular to the electric field direction[4,9-11] (oxygen over-pressure equilibrated with the local oxygen potential would indicate isotropic pore/bubble formation at any grain boundaries, irrelevant to the field direction). Strikingly, Graves $et\ al.$[4] reported 1,100 h's reversible SOFC/SOEC operations at high current densities (+0.5/−1 A/cm$^2$ for SOFC/SOEC, respectively) with negligible degradation and argued that SOFC operation somehow healed the microstructural damages developed under SOEC model with unclear reasons, without which rapid increase in ohmic resistance and extensive grain-boundary cavities were identified under continuous SOEC operations. So there are strong hints that SOFC and SOEC have some fundamentally different features, which (i) drive degradation in "opposite" directions (ii) at grain boundaries under electrochemical driving forces. Interestingly, by solving the oxygen potential distributions inside polycrystalline electrolytes, we identified very large oxygen potential gradients (named as oxygen potential transitions hereafter) at grain boundaries inside SOC electrolytes, and their signs were opposite under SOFC and SOEC modes. The inversion of oxygen potential transitions shows the most un-symmetric behavior between SOFC and SOEC to date and would have significant implications on their stabilities. Mechanistically, such an inversion is originated from the different (electro-)chemical forces that drive ionic and electronic current under SOFC/SOEC operations. Lastly, although the present calculations were primarily conducted on YSZ, whose ionic and electronic conductivities are best known[14], the phenomenon is general to any mixed ionic and electronic conductors, for both electrolyte and electrode materials.

## II.   Formulation of the problem

We shall consider an YSZ cell with hydrogen electrode (the electrode with a low oxygen potential) on the left and oxygen electrode (the electrode with a high oxygen potential) on the right, and define $x$ direction pointing from the left to the right. (Here we treat the transport as a one-dimensional problem for simplicity.) Under SOFC mode, the oxygen electrode is the cathode and the hydrogen electrode is the anode; the ionic current flowing from anode (left-hand side) to the cathode (right-hand side) is positive, while the electronic current flowing from cathode (right-hand side) to anode is negative (left-hand side); the total current is also positive since the ionic current is much larger than the electronic one in an YSZ fuel cell. Under SOEC mode, the oxygen electrode is the anode and the hydrogen electrode is the cathode; both the ionic and electronic currents are negative, flowing from cathode (right-hand side) to the anode (left-hand side); hence, the total current is also negative. Under the open circuit voltage (OCV), the ionic current flowing from anode (left-hand side) to the cathode (right-hand side) is positive, while the electronic current flowing from cathode (right-hand side) to anode is negative (left-hand side); they are opposite in sign and equal in magnitude, hence the total current is zero. Now given the oxygen potentials $\mu'_{O_2}$ at the hydrogen electrode/electrolyte interface and $\mu''_{O_2}$ at the oxygen electrode/electrolyte interface ($\mu'_{O_2} < \mu''_{O_2}$), we seek to solve the oxygen potential distribution inside the electrolyte. Phenomenologically, there are four species to be considered, i.e. oxygen ion $O^{2-}$, oxygen molecule $O_2$, electron e and hole h. Under local equilibrium, the two chemical reactions

$$O^{2-} = \tfrac{1}{2}O_2 + 2e \qquad (1)$$

$$e + h = \text{nil} \qquad (2)$$

relate the four potentials by

$$\tilde{\mu}_{O^{2-}} = \tfrac{1}{2}\mu_{O_2} + 2\tilde{\mu}_e \qquad (3)$$

$$\tilde{\mu}_e + \tilde{\mu}_h = 0 \quad (4)$$

where $\tilde{\mu}_{O^{2-}}$, $\tilde{\mu}_e$ and $\tilde{\mu}_h$ denote the electrochemical potential of $O^{2-}$, electrons and holes, respectively; $\mu_{O_2}$ denotes the oxygen potential. We next write the fluxes as

$$I_{O^{2-}}^{B} = \frac{\sigma_{O^{2-}}^{B}}{2e} \frac{d\tilde{\mu}_{O^{2-}}}{dx} \quad (5)$$

$$I_{O^{2-}}^{GB} = \frac{\sigma_{O^{2-}}^{GB}}{2e} \frac{d\tilde{\mu}_{O^{2-}}}{dx} \quad (6)$$

$$I_{e}^{B} = \frac{\sigma_{e}^{B}}{e} \frac{d\tilde{\mu}_{e}}{dx} \quad (7)$$

$$I_{e}^{GB} = \frac{\sigma_{e}^{GB}}{e} \frac{d\tilde{\mu}_{e}}{dx} \quad (8)$$

$$I_{h}^{B} = -\frac{\sigma_{h}^{B}}{e} \frac{d\tilde{\mu}_{h}}{dx} = \frac{\sigma_{h}^{B}}{e} \frac{d\tilde{\mu}_{e}}{dx} \quad (9)$$

$$I_{h}^{GB} = -\frac{\sigma_{h}^{GB}}{e} \frac{d\tilde{\mu}_{h}}{dx} = \frac{\sigma_{h}^{GB}}{e} \frac{d\tilde{\mu}_{e}}{dx} \quad (10)$$

where $I_{O^{2-}}$, $I_e$ and $I_h$ denote the current densities of $O^{2-}$, electrons and holes, respectively; $\sigma_{O^{2-}}$, $\sigma_e$ and $\sigma_h$ denote the conductivities of $O^{2-}$, electrons and holes, which could vary as a function of the temperature and the local oxygen potential; the superscript B and GB denote the corresponding quantity in bulk (referring to grain interior) and grain boundary (referring to space charge layer), respectively. Since electrons and holes can be generated and annihilated via Eq. (2), the fluxes of electrons and holes can be combined to be a total electronic current density $I_{eh}$

$$I_{eh}^{B} = I_{e}^{B} + I_{h}^{B} = \frac{\sigma_{e}^{B} + \sigma_{h}^{B}}{e} \frac{d\tilde{\mu}_{e}}{dx} \quad (11)$$

$$I_{eh}^{GB} = I_{e}^{GB} + I_{h}^{GB} = \frac{\sigma_{e}^{GB} + \sigma_{h}^{GB}}{e} \frac{d\tilde{\mu}_{e}}{dx} \quad (12)$$

At steady state, we assume no internal chemical reactions between ionic and electronic species (no generation/consumption of molecular oxygen inside the dense electrolyte), so both the ionic current density $I_{O^{2-}}$ and the electronic current density $I_{eh}$ remain constants throughout the electrolyte.

$$I_{O^{2-}} = I_{O^{2-}}^{B} = \frac{\sigma_{O^{2-}}^{B}}{2e}\frac{d\tilde{\mu}_{O^{2-}}}{dx} = I_{O^{2-}}^{GB} = \frac{\sigma_{O^{2-}}^{GB}}{2e}\frac{d\tilde{\mu}_{O^{2-}}}{dx} \quad (13)$$

$$I_{eh} = I_{eh}^{B} = \frac{\sigma_{e}^{B}+\sigma_{h}^{B}}{e}\frac{d\tilde{\mu}_{e}}{dx} = I_{eh}^{GB} = \frac{\sigma_{e}^{GB}+\sigma_{h}^{GB}}{e}\frac{d\tilde{\mu}_{e}}{dx} \quad (14)$$

Using Eq. (13-14), we can express the electrochemical potentials of oxygen ions, electrons and holes in terms of $I_{O^{2-}}$ and $I_{eh}$

$$\frac{d\tilde{\mu}_{O^{2-}}}{dx} = \begin{cases} 2e\dfrac{I_{O^{2-}}}{\sigma_{O^{2-}}^{B}}, & \text{inside bulk} \\ 2e\dfrac{I_{O^{2-}}}{\sigma_{O^{2-}}^{GB}}, & \text{at grain boundary} \end{cases} \quad (15)$$

$$\frac{d\tilde{\mu}_{e}}{dx} = -\frac{d\tilde{\mu}_{h}}{dx} = \begin{cases} e\dfrac{I_{eh}}{\sigma_{e}^{B}+\sigma_{h}^{B}}, & \text{inside bulk} \\ e\dfrac{I_{eh}}{\sigma_{e}^{GB}+\sigma_{h}^{GB}}, & \text{at grain boundary} \end{cases} \quad (16)$$

Plug Eq. (15-16) into Eq. (3) and rearrange, we get

$$f(x) = \frac{d\mu_{O_2}}{dx} = \begin{cases} 4e\left(\dfrac{I_{O^{2-}}}{\sigma_{O^{2-}}^{B}} - \dfrac{I_{eh}}{\sigma_{e}^{B}+\sigma_{h}^{B}}\right), & \text{inside bulk} \\ 4e\left(\dfrac{I_{O^{2-}}}{\sigma_{O^{2-}}^{GB}} - \dfrac{I_{eh}}{\sigma_{e}^{GB}+\sigma_{h}^{GB}}\right), & \text{at grain boundary} \end{cases} \quad (18)$$

Under SOFC mode, Eq. (18) can be written as

$$f(x) = \begin{cases} 4eI_{total}\left(\dfrac{t_i}{\sigma_{O^{2-}}^{B}} + \dfrac{t_{eh}}{\sigma_{e}^{B} + \sigma_{h}^{B}}\right), & \text{inside bulk} \\ 4eI_{total}\left(\dfrac{t_i}{\sigma_{O^{2-}}^{GB}} + \dfrac{t_{eh}}{\sigma_{e}^{GB} + \sigma_{h}^{GB}}\right), & \text{at grain boundary} \end{cases} \quad (19)$$

where $I_{total}$ is the total current density, $t_i$ and $t_{eh} = t_i - 1$ denote the ionic and electronic transference numbers, respectively. Under SOEC mode, Eq. (18) can be written as

$$f(x) = \begin{cases} 4e|I_{total}|\left(\dfrac{t_{eh}}{\sigma_{e}^{B} + \sigma_{h}^{B}} - \dfrac{t_i}{\sigma_{O^{2-}}^{B}}\right), & \text{inside bulk} \\ 4e|I_{total}|\left(\dfrac{t_{eh}}{\sigma_{e}^{GB} + \sigma_{h}^{GB}} - \dfrac{t_i}{\sigma_{O^{2-}}^{GB}}\right), & \text{at grain boundary} \end{cases} \quad (20)$$

where $t_{eh} = 1 - t_i$. Under OCV, Eq. (18) can be written as

$$f(x) = \begin{cases} 4eI_{0}\left(\dfrac{1}{\sigma_{O^{2-}}^{B}} + \dfrac{1}{\sigma_{e}^{B} + \sigma_{h}^{B}}\right), & \text{inside bulk} \\ 4eI_{0}\left(\dfrac{1}{\sigma_{O^{2-}}^{GB}} + \dfrac{1}{\sigma_{e}^{GB} + \sigma_{h}^{GB}}\right), & \text{at grain boundary} \end{cases} \quad (21)$$

where $I_0 = I_{O^{2-}} = -I_{eh}$ denotes the absolute value of the ionic and electronic current. Therefore, Eq. (19-21) leaves with only one unknown constant to be solved, i.e. $t_i$ for Eq. (19-20) and $I_0$ for Eq. (21), which can be determined by matching the boundary condition

$$L = \int_{\mu'_{O_2}}^{\mu''_{O_2}} \dfrac{d\mu_{O_2}}{f(x)} \quad (22)$$

where $L$ is the thickness of the electrolyte. Finally, the oxygen potential distribution can be obtained by integrating $f(x)$ from $x=0$ to arbitrary $x$

$$x = \int_{\mu'_{O_2}}^{\mu_{O_2}} \dfrac{d\mu_{O_2}}{f(x)} \quad (23)$$

Alternatively, for sufficiently small interval $\Delta l$, we may write the boundary condition as

$$\mu_{O_2}'' - \mu_{O_2}' = \sum_{n=0}^{L/\Delta l - 1} f(n\Delta l)\Delta l \qquad (24)$$

And the oxygen potential distribution can be obtained from

$$\mu_{O_2}(x) = \mu_{O_2}' + \sum_{n=0}^{x/\Delta l - 1} f(n\Delta l)\Delta l \qquad (25)$$

While the above two methods are mathematically equivalent, the second one turns out to be numerically much simpler for solving the considered polycrystalline problem (a multilayer problem) and will be used to obtain numerical results in **Section III**.

### III. Results

Numerical results were obtained to illustrate the effect of grain boundaries for two representative conditions: (1) 10 μm polycrystalline YSZ electrolyte at 800 °C, $\mu_{O_2}' = -4.26$ eV corresponding to an oxygen partial pressure $PO_2 = 10^{-20}$ atm, $\mu_{O_2}'' = 0.21$ eV corresponding to $PO_2 = 10$ atm, similar to Ref. 4; (2) 200 μm polycrystalline YSZ electrolyte at 1000 °C, $\mu_{O_2}' = -3.79$ eV corresponding to an oxygen partial pressure $PO_2 = 10^{-15}$ atm, $\mu_{O_2}'' = -0.18$ eV corresponding to $PO_2 = 0.2$ atm, similar to Ref. 7. In both cases, we consider a grain size of 2.5 μm, thickness of the space charge layer to be 10 nm, and current densities of +1/−1 A/cm² under SOFC/SOEC modes, respectively. For YSZ bulk, we use the conductivity data in Ref. 14, assuming $\sigma_{O^{2-}}^B$ independent of $\mu_{O_2}$, $\sigma_e^B \propto \exp(-\frac{\mu_{O_2}}{4RT})$ and $\sigma_h^B \propto \exp(\frac{\mu_{O_2}}{4RT})$ under standard defect chemistry considerations. For YSZ grain boundary, we use

$\sigma_{O^{2-}}^{GB} = \sigma_{O^{2-}}^{B}/100$, based on AC impedance measurements[24,25]; we further assume $\sigma_{e}^{GB} = \sigma_{e}^{B}$ and $\sigma_{h}^{GB} = \sigma_{h}^{B}$ because no information about grain-boundary electronic conductivity is available. The conductivity data are plotted in **Fig. 1**.

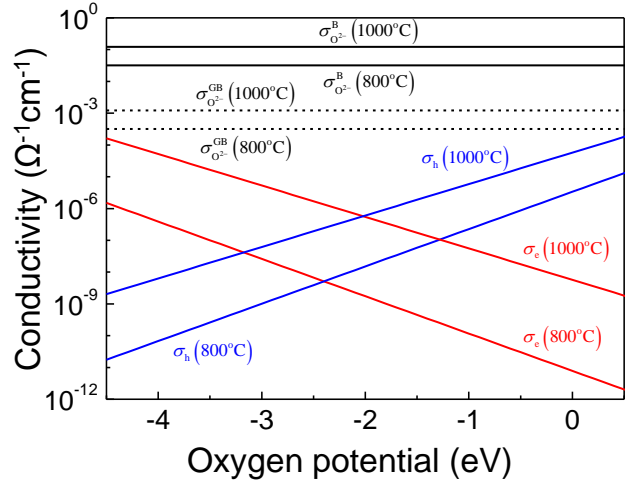

**Figure 1** Conductivity data of YSZ: $\sigma_{O^{2-}}^{B}$ (solid line in black) and $\sigma_{O^{2-}}^{B} = \sigma_{O^{2-}}^{GB}/100$ (dash line in black) for oxygen ions in the bulk and at grain boundary, $\sigma_{e}^{B} = \sigma_{e}^{GB} = \sigma_{e}$ for electrons (in red), $\sigma_{h}^{B} = \sigma_{h}^{GB} = \sigma_{h}$ for holes (in blue). Oxygen potential set to be zero at 1 atm.

The calculated oxygen potential distributions and gradients for Case (1) and (2) are shown in **Fig. 2** and **3**, respectively. Several feature should be noticed. First, all the curves in **Fig. 2a** and **3a** have a sigmoid shape, with the largest oxygen potential gradient at the oxygen potential corresponding to electronic conductivity minimum ($\sigma_e + \sigma_h$) in **Fig. 1**. Under the same magnitude of current density, the oxygen potential distribution is steeper under SOEC mode than under SOFC mode (by comparing red and blue curves in **Fig. 2** and **3**), and steeper for a thinner electrolyte than a thicker one

(by comparing **Fig. 2** and **3**). These have been recognized and are consistent with previous theoretical studies[7,8,22,23]. Second, although grain boundaries are 100 times more $O^{2-}$ blocking than bulk, the oxygen potential distributions inside polycrystals do not differ much from the corresponding references inside single crystals. This is understandable since grain boundary is much thinner than grain size (10 nm vs. 2.5 μm in the present cases) and constitutes only small portion of the total thickness. Third, the existence of grain boundaries slightly sharpens the oxygen potential distribution under SOEC mode and slightly smoothens it under SOFC mode. Fourth, there are an oxygen potential transition at each grain boundary, with a large oxygen potential gradient. For case (1) in **Fig. 1**, the oxygen potential gradients at grain boundaries are about 10 times smaller than the largest one at electronic conductivity minimum in the bulk; for case (2) in **Fig. 2**, the ones at grain boundaries are much larger than the largest one in the bulk. Lastly, and most interestingly, the oxygen potential transitions at grain boundaries are inverse and the gradients have opposite signs under SOFC vs. SOEC modes.

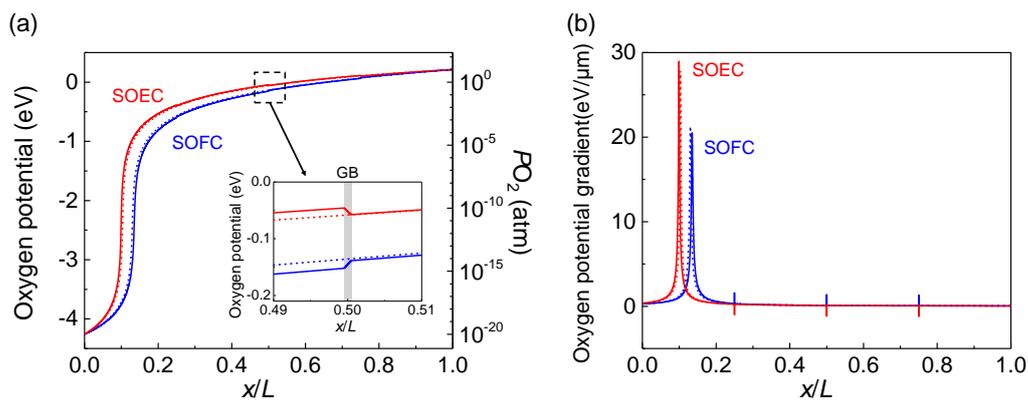

**Figure 2** Calculated (a) oxygen potential distribution and (b) oxygen potential gradient for polycrystalline YSZ electrolyte at +1 A/cm$^2$ under SOFC mode (solid line in blue) and at −1 A/cm$^2$ under SOEC mode (solid line in red). Electrolyte thickness: 10 μm;

grain size: 2.5 μm; space-charge layer thickness: 10 nm; temperature: 800 °C; $\mu'_{O_2} = -4.26 \text{ eV}$, $\mu''_{O_2} = 0.21 \text{ eV}$. References for single-crystalline electrolyte without grain boundaries also plotted under SOFC (dash line in blue) and SOEC mode (dash line in red). Inset of (a): Magnified view around a grain boundary.

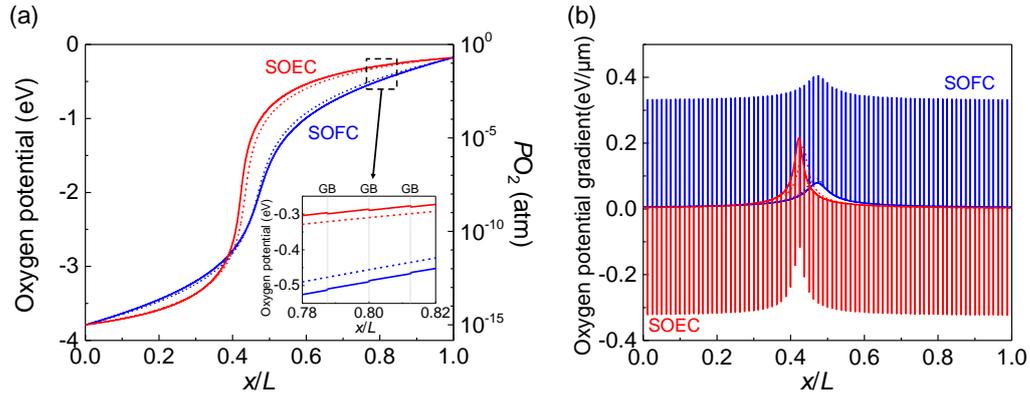

**Figure 3** Calculated (a) oxygen potential distribution and (b) oxygen potential gradient for polycrystalline YSZ electrolyte at +1 A/cm² under SOFC mode (solid line in blue) and at −1 A/cm² under SOEC mode (solid line in red). Electrolyte thickness: 200 μm; grain size: 2.5 μm; space-charge layer thickness: 10 nm; temperature: 1000 °C; $\mu'_{O_2} = -3.79 \text{ eV}$, $\mu''_{O_2} = -0.18 \text{ eV}$. References for single-crystalline electrolyte without grain boundaries also plotted under SOFC (dash line in blue) and SOEC mode (dash line in red). Inset of (a): Magnified view around grain boundaries.

## IV. Discussions

Obviously, the oxygen potential transitions at grain boundaries come from the different transport properties of the grain boundary with the bulk. To maintain constant ionic and electronic current densities across the electrolyte, a larger electrostatic potential gradient is spent at the more $O^{2-}$ blocking grain boundaries and a corresponding change in the chemical potential gradient of electrons and holes is required to suppress the over-flow of electrons and holes. Under the assumption of local equilibrium, chemical potential of electrons and holes is always equilibrated with oxygen potential. Therefore, the large chemical potential gradient of electrons and holes is reflected by the oxygen potential transition plotted in **Fig. 2a** and **3a**. Conceptually, this is in the same spirit with the overpotential across the electrode/electrolyte interface: avoiding discontinuity of fluxes at heterogeneous interfaces. So the oxygen potential transition at grain boundary can be integrated over the thickness to define a "grain-boundary overpotential". Their difference is: electrode overpotential drives chemical reactions involving oxygen gases, while grain-boundary overpotential drives ion/electron/hole fluxes without reactions. To weaken such oxygen potential transitions at grain boundaries, or lower grain boundary overpotential, one should make bulk conductivities and grain boundary conductivities more alike. This is along the same route people have been trying to decrease space-charge potential and increase grain boundary conductivity of $O^{2-}$ via grain boundary engineering.

Now we come to the question: why there is an inversion of oxygen potential transition at grain boundaries in SOFC/SEOC? To address this, one should first notice that ionic and electronic currents flow along the opposite directions in SOFC but along the same direction in SOEC[5]. In YSZ electrolyte, the chemical potential of $O^{2-}$ is fixed because extensive aliovalent doping pins oxygen vacancy concentration[7]. Therefore, ionic current is driven by electrostatic potential alone. For electrons and holes, their concentrations could differ over many orders of magnitude at two electrodes so in addition to electrostatic potential, chemical potential of electrons and holes can also drive electronic current. SOFC operates under an oxygen potential difference or a Nernst voltage to produce electricity. By definition, the Nernst voltage should be larger than the integral of electrostatic potential gradient. In the electrochemical potential of electrons and holes, the electrical and chemical parts are opposite in sign and the latter being larger in magnitude determines the direction of electronic current. Therefore, ionic current controlled by electrostatic potential flows oppositely to electronic current. In comparison, SOEC operates under an applied voltage across an oxygen potential difference. By definition, the integral of electrostatic potential gradient should be larger than the Nernst voltage. Therefore, the electrical part is larger than the chemical part in the electrochemical potential of electrons and holes, and ionic and electronic currents flow in the same direction.

Back to the polycrystal problem, under both SOFC and SOEC modes, we have positive oxygen potential gradient in the bulk (**Fig. 2 and 3**). As discussed earlier, electrostatic potential gradients are larger at more $O^{2-}$ blocking grain boundaries. As a result, the chemical part in the electrochemical potential of electrons and holes needs to cancel the electrical part to compensate the otherwise over-flow of electrons and holes. In SOFC, the electrical and chemical parts are opposite in sign so chemical part only needs to become larger without changing the sign. Therefore, oxygen potential gradients are both positive at grain boundaries and in bulk. In SOEC, the electrical and chemical parts have the same sign, so in order to cancel the electrical part, the chemical part must change its sign. Therefore, oxygen potential gradient becomes negative at grain boundaries. This clarifies the origin of inverse oxygen potential transitions at different operation modes.

To this point, it is interesting to note such oxygen potential transitions have the same characteristics with the electrode overpotential. In SOFC, oxygen potentials at electrode/electrolyte interfaces are bounded by the two gaseous atmospheres. At both hydrogen electrode (left electrode in our definition)/electrolyte interface and oxygen electrode/(right electrode in our definition)/electrolyte interface, oxygen potential is lower on the left-hand side than the right-hand side. The same trend applies to grain boundaries: oxygen potential is lower on the left-hand side than the right-hand side as shown in **Fig. 2a** and **3a**. Similarly, in SOEC, oxygen potentials at electrode/electrolyte interfaces are outbounded by the two gaseous atmospheres. This is to say, at both electrode/electrolyte interfaces, oxygen potential is higher on the left-hand side than the right-hand side. Again, the same trend applied to grain boundaries, as shown in **Fig. 2a** and **3a**. Therefore, electrode overpotentials have the same signs as oxygen potential transitions, or grain-boundary overpotentials, and it is well known that electrode overpotentials are inversed for SOFC and SOEC! More interestingly, this would raise the following question. Jacobsen and Mogensen[7] wrote "oxygen pressure inside the electrolyte will never become higher than the pressure corresponding to the electrode potential of the oxygen electrode and never lower than corresponding to the electrode potential of the hydrogen electrode, irrespective of which mode or condition for the cell operation." While the statement still holds under SOFC mode, it may break down and oxygen potential could be un-bounded in polycrystalline electrolyte with $O^{2-}$ blocking grain boundaries under SOEC mode, and the more $O^{2-}$ blocking the more so. That is to say, the highest oxygen pressure could be at the grain boundary of the electrolyte next to the oxygen electrode, which provides the highest driving force as well as preferential nucleation sites for oxygen bubble formation. Such a possibility has been also discussed by Chatzichristodoulou *et al.*[23] at the interface of YSZ/$Gd_{0.1}Ce_{0.9}O_{1.95}$ bilayer electrolyte under SOEC mode.

The inversion of oxygen potential transition could have significant influence on stability of polycrystalline electrolyte. Specifically, we see the oxygen potential gradients at grain boundaries and in bulk are same in sign in SOFC but opposite in sign in SOEC. It is likely that grain boundaries experience different chemical stresses in SOFC and SOEC. If pore/bubble formation and other degradation processes are stress driven and nucleation-controlled, it may explain why reversible SOFC/SOEC operations eliminate degradations. This effect could be further explored by analyzing stress states using calculated oxygen potential distributions as inputs, with finite-element modeling under different boundary conditions (e.g. anode or electrolyte supported SOC).

Lastly, one should note the present study is based on continuum level. In reality, grain boundaries and space charge layers are only a few nanometers thick, and charge neutrality does not hold either. The influence of atomic discretization would be interesting yet difficult to consider. We also assume all grain boundaries have the same transport properties, in the same way AC impedance measurement does. However, grain boundaries with distinct misorientations and structures could behave distinctly, which may lead to degradation preferentially at some special grain boundaries, for example a more $O^{2-}$ blocking one. These complexities could smear the phenomena, but we believe the general trend still holds.

## V.     Conclusions

(1)  By solving transport equations in polycrystalline electrolyte, we identified a sharp oxygen potential transition at grain boundaries and their directions are inverse when operated under SOFC and SOEC modes.

(2)  Ionic and electronic currents flow along opposite directions in SOFC and along the same direction in SOEC, which is rooted in the different (electro-)chemical forces that drive ionic and electronic current under SOFC/SOEC operations. The inversion of oxygen potential transitions has the same origin.

(3)  It is suspected the inversion of oxygen potential transitions lead to different stress states at grain boundaries under SOFC/SOEC modes, which is related to their contrasting degradation kinetics.

(4)  Oxygen potential could be un-bounded by two terminal ones for polycrystalline electrolyte operated under SOEC mode and internal oxygen pressure could be the highest at the grain boundary of the electrolyte next to the oxygen electrode.

(5)  Modifying grain boundary conductivities could weaken or eliminate oxygen potential transitions at grain boundaries and provides better stability of electrolyte materials.